\documentclass[twocolumn]{autart}    
\usepackage[modulo,switch]{lineno}
\modulolinenumbers[1]
\usepackage{amsmath}
\usepackage{hyperref}
\usepackage{amssymb}
\usepackage{graphicx}
\usepackage{float}
\graphicspath{{images/}{../images/}}
\usepackage{caption}
\usepackage{afterpage}
\usepackage{fancyhdr} 
\usepackage{pdfpages}
\usepackage{subfiles}
\usepackage{blindtext}
\usepackage{setspace}
\usepackage[round]{natbib}
\renewcommand{\refname}{REFERENCES:}
\usepackage{enumitem}                                     

\usepackage{graphicx}          

\begin{document}

\begin{frontmatter}

\title{On The Ignition, Propagation and Termination Of The Neuronal Bursting Activity During Ictogenesis In Epileptic Patients} 


\author{Eslam Abbas}

\address{Harmel Hospital, 987 Nile Corniche, Al Kafor, Cairo, 11559, Egypt.}  

\begin{keyword}                           
Epilepsy; Ictogenesis, Excitatory Feedbacks; Competition Zone; and Escape Rhythm               
\end{keyword}                             

\begin{abstract}                          
Epilepsy creates a persistent increase in the probability of spontaneous seizures. An ictal episode evolves due to acute disturbance of the fine-tuned balance between excitatory vs. inhibitory inputs within a neural network in favor of excitation. The current literature that proposes the activity-dependent disinhibition as a valid mechanism of chronic epilepsy, does not provide clues on why this mechanism emerges only in epileptic patients and how the vicious circle resulting of an activity-dependent disinhibition in over-active ictogenic network would end. A new model, which presents chronic epilepsy as a disease of faulty architecture of the neural circuit, is discussed. Wherein; variable genetic or acquired predisposing factors drive abnormalities in the construction of multiple neural circuits resulting in an activity-dependent positive feedback excitatory loops which transform normal neural circuits into ictal foci. Such new mechanism, for igniting an activity-dependent unstable excitation with subsequent relatively stable disinhibition, leads to an ictal escape rhythm. The propagation of such bursting activity occurs either electrochemically via synaptic communication to remote susceptible circuits, or chemically via a trigger wave which recruits the non-connected proximal neurons. Termination occurs abruptly when the inhibitory interneurons functionally recover and reimpose their inhibitory effect on the ictogenic circuit to transform the escape rhythm into a normal, under-control output. The proposed model elucidates various enigmatic features of the disease; and illustrates both the end-result ictogenic mechanism arising from the wide variety of etiologies of human spontaneous and acquired epilepsy, and the timing of episodic transitions from normal activity to seizures.
\end{abstract}

\end{frontmatter}

\section*{Introduction}
\noindent Brain electrical activity is non-synchronous \citep{mcphee1995pathophysiology} and regulated by factors within the neuron and the extraneuronal environment. Neuronal factors comprise the type, number and distribution of ion channels; changes to receptors of neurotransmitters and modulation of gene expression \citep{bromfield2006basic}. The extraneuronal factors include ion concentration, synaptic plasticity and regulation of transmitters' breakdown and reuptake \citep{blumenfeld2005cellular}.\par

\noindent During seizures; a discrete group of neurons begins abnormal excessive firing in a synchronized manner \citep{da2003epilepsies}\citep{margineanu2010epileptic} which then can propagate to neighboring regions \citep{gotz2008spread}. The characteristics of said firing are high frequency bursts of action potentials and hypersynchronization \citep{fisher2005epileptic}. At the level of single neurons, the ictal discharge shows a characteristic paroxysmal depolarization shift, which consists of a sequence of sustained neuronal depolarization resulting in a burst of action potentials, a plateau-like phase associated with completion of the action potential bursts, and then rapid repolarization followed by hyperpolarization \citep{misulis2003essentials}.\par

\noindent The exact mechanism by which such smooth normal brain activity is suddenly shifted to the bursting ictal firing is still unclear \citep{noebels2012jasper}. The axiomatic mechanism, which involves an acute imbalance between the excitatory discharge and the inhibitory control, needs further elaboration on the cellular and molecular levels. Mutagenic neurochemical changes in the neurotransmitters, receptors and ion channels are a valid mechanism to address severe forms of epileptic encephalopathies wherein the brain activity is distorted with no or little normal epochs. Nevertheless; neurochemical changes on such a molecular level is time-invariant and don't provide an explanation for the unpredictable episodic, and relatively rare, seizures \citep{schulze2008unpredictability} that occurs in chronic epilepsy. Besides; non-mutagenic changes in the number and distribution of ion channels and neurotransmitter receptors can not be asserted as a primary mechanism of epilepsy's pathogenesis, or a reactive remodeling as a result of repeated ictal behavior of epileptogenic foci.\par

\noindent Models of activity-dependent disinhibition for ictogenesis solve the timing problem; wherein disinhibition occurred only at the extremes of network activity \citep{bracci2001dynamic}\citep{wester2014behavioral}. Therefore; the probability of ictal changeover would depend on the probability of that exceptional level of activity. Likely; the same range of probabilities characterizing spontaneous seizures can agree with the odds of these levels of activity \citep{staley2015molecular}. Once a seizure is induced by an activity-dependent disinhibition, the seizure itself can continue to produce activity levels that are sufficient to suppress inhibition, providing the necessary positive feedback to sustain the seizure. Yet; Said models do not provide the fundamental mechanism by which a robust input can transform a network with apparently normal activity into an ictal focus. Moreover; they do not give clues on how the vicious positive feedback circle will end, adding more ambiguity on the poorly understood mechanism of termination of an ictal activity.\par

\noindent A model for the pathogenesis of chronic epilepsy shall provide insights on the most enigmatic aspects of the disease like the basic fundamental mechanism that emerges from the wide varieties of epileptogenic etiologies, propagation, and termination of ictal behavior; besides the unpredictable episodic timing of ictal episodes.

\section*{MODEL:}

\noindent The ictogenesis process involves two main checkpoint steps, the initiation of the burst, and the propagation of the bursting activity. The initiation step involves abrupt imbalance between excitation and inhibition within the neural environment leading to the acute transition of normal brain activity to ictal rhythm. The proposed model functionally categorizes the origin of such abrupt imbalance into three different levels of neuronal hyperexcitability with ictal discharge, with special emphasis on the second intermediate level; the chronic epilepsy. While also proposes that the propagation checkpoint depends on the initial conditions; wherein if a large-enough number of neurons burst synchronously, the propagation of a depolarization wave becomes mandatory.\par

\noindent The first level is seizures in otherwise healthy brain tissue, which result from the simultaneous mass-excitation of large numbers of neurons by external insult or stimuli. Electric stimulation during ECT and chemical stimulation, either during acute overdosing or sudden withdrawal of a drug, are valid examples. During ECT; electrodes deliver an electrical impulse which traverses through intermediary brain tissue to simultaneously stimulate neurons by altering their internal electrical milieu and concentration of ions \citep{swartz2014mechanism}. Chemical stimuli may involve toxic exposure to domoic acid, which activates excitatory GluK1 glutamate receptors \citep{jett2012chemical}, or by overdoses of theophylline, which blocks the inhibitory adenosine A1 receptor \citep{boison2011methylxanthines}. Also; abrupt withdrawal of GABAergic-acting sedative--hypnotic drugs can cause seizure due to chronic GABA receptor downregulation as well as glutamate overactivity, which lead to neurotransmitter sensitization and neuronal hyperexcitability \citep{allison2003neuroadaptive}. Accordingly and as a result of both electrical and chemical stimulation; a large focus of hyperexcitable neurons bursts synchronously bypassing the initiation checkpoint to induce a depolarization wave which effectively propagates to induce an ictal episode, or series of episodes in case of chemical insults.\par

\noindent The third level comprises severe epileptic encephalopathies, which result from severe genetic abnormalities that compromise important inhibitory pathways. Such mutations are most frequently associated with continuous altering of brain functions; wherein there are no normal epochs of brain activity, with subsequent frequent seizures \citep{allen2013novo}\citep{veeramah2013exome}. Said seizures are usually multiform and intractable and usually accompanied by relentless cognitive, behavioral and neurological deficits \citep{khan2012epileptic}.\par

\noindent The second intermediate level is chronic epilepsy, which is emphasized in this model as it is responsible for the vast majority of seizures in humans. Chronic epilepsy can be subdivided into spontaneous, and acquired due to an acute injury of the normal brain tissue as trauma, strokes and infections. Despite different pathological origin; both subdivisions leads to extemporaneous activity-dependent shifts in the balance between inhibition and excitation in one or more neural circuits to increase the probability of seizures break out. This model tracks the poorly-understood initiation, propagation and termination of the ictal episode in chronic epilepsy.\par

\begin{figure*}
    \centering
    \includegraphics[width=0.8\linewidth]{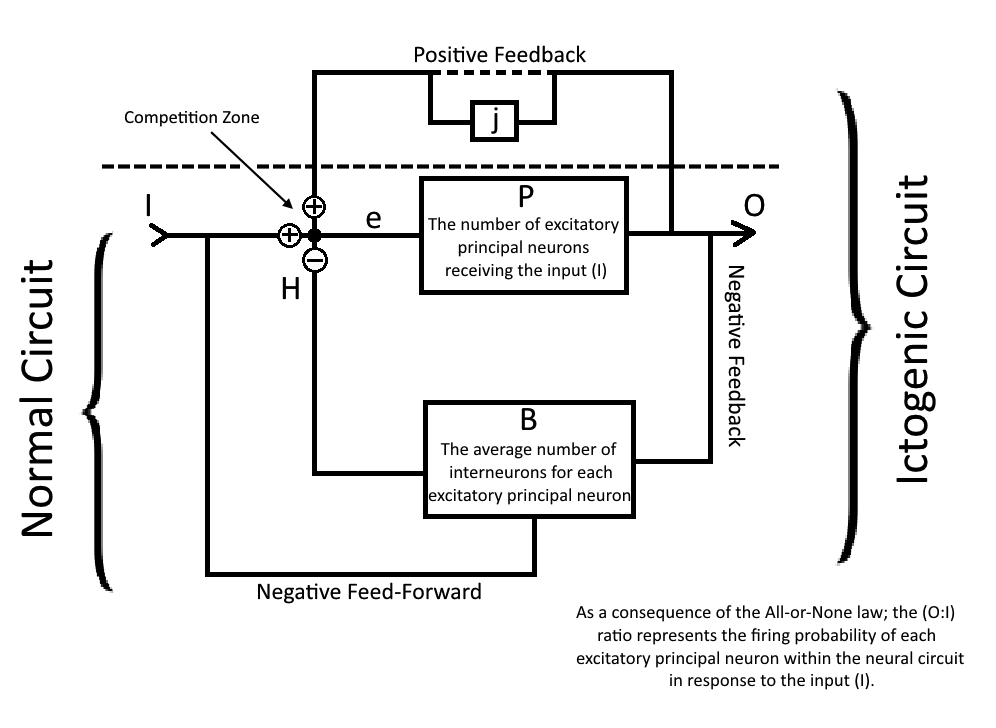}
\caption{\textit{A block diagram describes the architecture of the normal neural circuit vs. the ictogenic circuit.}}
\end{figure*}

\noindent Within a normal neural circuit; the excitatory output of principal cells is usually put under control by the interneurons which mainly exert an inhibitory effect on the incoming inputs; wherein the triggering of an action potential is determined by summation of excitatory and inhibitory signals. A neural circuit can be analyzed according to its neuronal content, to determine the feedback, feedforward balance between the inhibitory and the excitatory neurons, to yield a block diagram as shown in \textbf{Fig. 1}; wherein \((I)\) represents the input excitatory signal, \((H)\) represents the input inhibitory signal from the interneurons, \((e)\) is the summation effect and represents the actual input, \((P)\) is the number of excitatory principal neurons receiving the input \((I)\) within the neural circuit, \((B)\) is the average number of inhibitory interneurons for each excitatory principal neuron within the neural circuit, and \((O)\) represents the output of the neural circuit. The output of the circuit is:
\begin{equation*}
O=eP
\end{equation*}
\noindent and the actual input after summation is:
\begin{equation*}
e=I-H
\end{equation*}
\noindent then, the output of the circuit can be described as:
\begin{equation*}
O(1+BP)=IP
\end{equation*}
\noindent so, the ratio between the output and the input is:
\begin{equation*}
\frac{O}{I}=\frac{P}{1+BP}
\end{equation*}\par

\noindent This equation represents a negative feedback system, wherein the emergent output of the circuit is lesser than the input signal by the effect of (\(B\), the average number of inhibitory interneurons for each excitatory principal neuron within the neural circuit). Herein and as a consequence of all or none law; the \(\frac{O}{I}\) ratio actually represents the firing probability of each excitatory principal neuron within the neural circuit in response to the input \((I)\).
\begin{equation*}
\frac{O}{I}< 1
\end{equation*}\par

\noindent Negative feedback is a widespread inhibitory tuning mechanism within neural circuits, but negative feedforward usually augments a robust control as it's characterized by inhibition certainty and minimal time lag. The inhibitory interneurons within a neural circuit not only put the excitatory output of the principal neurons under check, but also assure desynchronous firing of said neurons by decreasing the firing probability of each excitatory principal neuron within the neural circuit in response to a certain input. The presence of feed inhibition guarantees that the firing probability, of an excitatory neuron in response to an input signal, is usually less than one.\par

\noindent An ictogenic focus is characterized by defective neuronal architecture which leads to priming of positive feedback excitatory pathways consequential to interlaced excitatory interneurons or sprouting of axonal collaterals. Said defective neuronal architecture can result from the faulty neuronal arrangement due to dysgenesis guided by genetic predispositions, or as an end result of the compensatory reconstruction that occurs after brain injury. The positive feedback excitatory pathways are triggered off in an activity-dependent manner. When an episodic surge of the circuit activity reaches a critical value, the positive feedback excitatory pathways are fully activated. Then; the system enters a positive feedback loop wherein the excitatory impulses reverberate to self-reinforce further excitation.\par

\noindent The activity-dependent positive feedback loop is extremely unstable, as the value \((Oj)\) is usually equal or more than unity \((Oj\geq 1)\), and collapses spontaneously; wherein \((j)\) represents the average number of excitatory feedback interconnections for each excitatory principal neuron within the neural circuit. Yet, as a consequence of impulses reverberation and self-reinforced excitation; the interneuronal inhibitory network enters a state of stupor. When the inhibitory control is defunct \((H=0)\), the output of the ictogenic circuit is released. So; for each excitatory principal cell within the ictogenic circuit \((P=1)\), each input will result in a subsequent output \((I=O)\) forming a synchronous escape rhythm; wherein all the excitatory principal neurons receiving the input \((I)\) within the neural circuit will burst simultaneously upon receiving the input \((I)\). However; such self-reinforced signal magnification and the subsequent escape rhythm (ictal discharge) occurs only at the extremes of the circuit activity; wherein the probability of output release would depend on the probability of an exceptional level of a vigorous activity that reaches the critical value.\par

\noindent The bursting activity resulting from the escape rhythm can be confined due to the integral surrounding zone of inhibition, which cannot be overcome by the current level of ictal discharge, to produce a focal abnormality of the brain electrical activity; or can propagate to distort the electrical activity of the whole or a large portion of the brain. The bursting ictal discharge will not overcome the zone of inhibition as long as the number of excitatory principal neurons, receiving an input \((I)\) within an ictogenic neural circuit, is small (\(P\) is small) leading to failure of electrochemical recruitment via synaptic communication; and/or failure of chemical recruitment of the non-connected proximal neurons as \((r\ll \phi_r)\); wherein \((r)\) is the radius of the recruited focus, and \((\phi_r)\) is the minimum critical focal radius from which the chemical recruitment can begin and the trigger wave can get off.\par

\noindent The propagation of the paroxysmal depolarization shift, as a characteristic bursting activity, can occur via synaptic transmission to the susceptible connected neurons of other remote networks; wherein the repetitive discharge from the presynaptic ictogenic neuron leads to \(Ca^{++}\) accumulation in the presynaptic terminals enhancing neurotransmitters release. Besides; it can propagate to the non-connected adjacent neurons by altering the extracellular ionic concentrations to produce a trigger wave which gradually recruits the neurons in the proximity to initiate a bursting activity by augmenting their excitability.\par

\noindent The neuronal resting membrane potential \((V_{resting})\) is mainly maintained by the role of \(K^+\) and \(Na^+\) leak channels \citep{purves2001neuroscience}. Due to the high concentration level of \(Na^+\) ions in the extracellular fluid and diminished membrane permeability; the effect of changes of the ion concentration in the extraneuronal environment on the resting membrane potential, differs greatly between \(K^+\) and \(Na^+\) ions. Such effect can be represented by the following two equations:
\begin{equation*}
V_{resting}\propto - \frac{K_{in}^+ - K_{out}^+}{K_{out}^+}
\end{equation*}
\begin{equation*}
V_{resting}\propto + \frac{Na_{out}^+ - Na_{in}^+}{Na_{out}^+}
\end{equation*}\par

\noindent Both equations show the two different effects of fluctuations in the extraneuronal concentration of \(K^+\) and \(Na^+\) ions on the resting membrane potential. Accordingly and regardless of the osmotic effect; while the electrochemical effect of hypo/hypernatremia is negligible, neuronal resting membrane potential is very sensitive to trivial fluctuation of extraneuronal potassium concentration. The bursting activity resulting from the escape rhythm leads to a transient increase in the extraneuronal \(K^+\) concentrations within the proximity of the neural network. Said \(K^+\) fluctuations increase the excitability of adjacent neurons due to depolarizing the resting membrane potential. An increase in the neuronal excitability sets the affected neurons into bursting activity, leading to the propagation of the PDS through the brain tissue. As a trigger wave, each affected neuron augments further propagation, so that the propagation of PDS doesn't slow down or lose amplitude as it travels through the brain tissue. The rate of recruitment of neurons into the bursting activity, propagation of the trigger wave, can be represented by the following equation:
\begin{equation*}
\frac{d(N)}{dt}=\rho_n \bigg( \frac{d(V_{ol})}{dt} \bigg)
\end{equation*}
\noindent or
\begin{equation*}
\frac{d(N)}{dt}= 4\pi . \rho_n \bigg( r^2 . \frac{d(r)}{dt} \bigg)
\end{equation*}
\noindent wherein
\begin{equation*}
r \geq \phi_r
\end{equation*}
\noindent wherein \((N)\) is the number of recruited neurons in unit of time \((t)\), \((\rho_n)\) is the neuronal density per unit of volume \((V_{ol})\), and \((r)\) is the radius of the recruited focus, or the radius of the propagating trigger wave. \((\phi_r)\) is the minimum critical focal radius from which the chemical recruitment can launch. The bursting activity of the ictogenic circuit ends abruptly when the inhibitory interneurons within the ictogenic focus become functionally effective to transform the stable escape rhythm into a normal, under-control output.

\section*{RESULTS:}

Many results can be inferred from the proposed model, but hereinafter are some that elucidate various enigmatic features of chronic epilepsy. By presenting epilepsy as a disease caused by mal-architecture of the neural circuits within the brain tissue; \textbf{the probability of formation of spontaneous ictogenic foci increases within the brain regions characterized by high-rate of continuous active modulation of tissue micro-structure}. Additionally; the model probes into the genesis of the ictal activity in epileptic patients; wherein the initiation involves igniting an activity-dependent unstable positive feedback excitatory loop which spontaneously collapses, yet leads to a synchronous escape rhythm forming a bursting activity.\par

\noindent\textbf{The action potential generated by uninhibited ignition tends to have echoic reverberations}; so that the resulting escape rhythm shows the characteristic multi-spiking paroxysmal depolarization shift as a hallmark of epilepsy on the cellular level. The ictal bursting activity can be either confined by an intact zone of inhibition, which can't be overcome by the current level of ictal discharge (\(P\) is small and/or \(r\ll \phi_r\)), to produce a focal abnormality of the brain electrical activity; or propagating to distort the electrical activity of the whole or a large portion of the brain. The propagation of the bursting discharge occurs either electrochemically via synaptic communication to remote susceptible circuits, or chemically via a trigger wave which recruit the non-connected proximal neurons. Termination occurs abruptly when the inhibitory interneurons functionally recover and reimpose their inhibitory effect on the ictogenic circuit to transform the escape rhythm into a normal, under-control output.\par

\noindent Two architectural factors contribute to the aggressiveness of an ictogenic focus; a major factor which is \textbf{the average density}, or average number, of the excitatory feedback pathways for each excitatory principal neuron within the neural circuit \((j)\), and a minor factor which is \textbf{the ratio between the neuronal density to the synaptic density} \( \big(\frac{\rho_n}{\rho_s}\big)\) within the ictal focus. Additionally; the velocity of the chemical neuronal recruitment via a trigger wave is proportional to the neuronal density of the affected brain region.\par 

\noindent Moreover; the model shows that \textbf{spontaneous epilepsy is a dynamic non-static chronic disease}; wherein untreated recurrent attacks can lead to formation of neo-foci or exaggeration of the principle one via mal-architecting resulting of ectogenic excitotoxicity and subsequent faulty rewiring driven by genetic predispositions. On the other hand; a rare possibility of long-term decomposition of an ictogenic focus, due to neural plasticity, can occur if consolidation with further ictal attacks is prevented. Besides; paradoxical therapeutic effect of combined anti-epileptic drugs, which enhance the GABAergic hyperpolarization, can occur, due to the higher probability of main input magnification, \textbf{if the synaptic plasticity between the main presynaptic input and the postsynaptic membrane within the competition zone is anti-Hebbian}.

\section*{DISCUSSION:}

Several decades of experiments, which depend on pharmacologically-induced seizures, have established the idea that an imbalance between inhibitory and excitatory activity leads to ictal episodes \citep{scharfman2007neurobiology}. An acute imbalance between excitation and inhibition is thus a valid ictogenic mechanism. Problems arise when such mechanism is extended to cover the chronic process of epileptogenesis, which creates a persistent rise in the probability of spontaneous seizures \citep{staley2015molecular}. The timing of seizures in chronic epilepsy is unpredictable and ictal episodes are relatively rare, representing much less than 1\% of the total brain activity \citep{moran2004epilepsy}. Thus; in chronic epilepsy, an ictogenic mechanism shall also explain the timing of episodic transitions from normal to ictal brain activity.\par

\noindent Additionally; another difficulty arises when applying the model of imbalanced inhibition and excitation to the etiology of chronic epilepsy, which does not usually suggest a chronic imbalance. In genetic predisposed epilepsy, analyses of the genetic etiology have occasionally found causal loss-of-functions mutations in inhibitory pathways \citep{macdonald2012mrna}, but loss-of-function mutations are also found in several excitatory pathways \citep{frank2006reduced}\citep{carvill2014gabra1}, and the majority of causal mutations involve genes that do not directly alter the balance of inhibition and excitation \citep{ran2014epilepsygene}. In the acquired epilepsies, seizures start after an insult to a normal brain tissue as a result of stroke, trauma, or infection. Steady-state imbalances in excitation against inhibition, in established animal models of acquired epilepsy, are difficult to demonstrate. In the Pilocarpine model; a damage to the inhibitory neurons is compensated by an increase in GABAergic synaptogenesis before the onset of seizures \citep{zhang2009surviving}. Compensatory glutamatergic synaptogenesis also happens \citep{buckmaster2014does}, but steady-state network imbalances, between excitation and inhibition, are not evident in experimental and human epilepsy \citep{sutula2007unmasking}.\par

\noindent The current literature that proposes the activity-dependent disinhibition as a valid mechanism in chronic epilepsy, does not provide clues on why this mechanism emerges only in epileptic patients and how the vicious circle resulting of an activity-dependent disinhibition in over-active ictogenic network would end. The proposed model presents chronic epilepsy as a disease of faulty architecture of the micro-structure of the brain tissue; wherein variable genetic or acquired predisposing factors drive abnormalities in the construction of multiple neural circuits resulting in an activity-dependent positive feedback excitatory loops which transform normal neural circuits into ictal foci. Such new mechanism for igniting an activity-dependent unstable excitation with subsequent relatively stable disinhibition leads to an ictal escape rhythm. Additionally; Said model provides insights about the mechanism of propagation and termination of such bursting activity. Besides; it illustrates both the end-result ictogenic mechanism arising from the wide variety of etiologies of human spontaneous and acquired epilepsy, and the timing of episodic transitions from normal activity to seizures.

\subsection*{MOLECULAR MECHANISM OF IGNITION:}

Either genetic predispositions drives defective wiring in spontaneous epilepsy or after-insult dysgenesis in the acquired one; both lead to architectural flaws in one or more neural circuits transforming them into ictal foci; wherein positive feedback excitatory loops are formed via sprouting axonal collaterals and/or interlaced excitatory interneurons. Consequently; a competition zone is formed, which comprises presynaptic long-term potentiated main excitatory input synapse(s), presynaptic inhibitory synapse(s) (formed by feedback or feedforward inhibitory interneurons), presynaptic long-term depressed feedback excitatory synapse(s) (formed by positive feedback interneurons or axonal collaterals), and a postsynaptic membrane of a principle neuron. The competition zone is not an anatomical but a functional entity; wherein a dynamic contest among synaptic plasticities determine the polarization of the postsynaptic membrane and further generation of an action potential. Normally; when a main excitatory input comes into the competition zone, the activity of the inhibitory feedforward or feedback control dictates the probability of action potential generation and firing of the postsynaptic neuron, whereas the long-term depressed excitatory feedback plays a negligible role.\par

\noindent Spike-timing-dependent plasticity causes a difference in the homosynaptic plasticity between the presynaptic main excitatory input, which evolves a long-term potentiation, and the presynaptic excitatory feedback(s), which evolve long-term depression. Due to time lag; the presynaptic feedback excitatory synapse(s) are long-term depressed because they always fires immediately after the firing of the supplied postsynaptic principal neuron, hence this particular feedback excitatory stimulus is made weaker \citep{song2000competitive}\citep{debanne1994asynchronous}. Additionally; the rise in the synaptic weight of the presynaptic main excitatory pathway dictates a reduction in the synaptic weight of the presynaptic feedback excitatory synapse(s) to keep the average synaptic weight approximately conserved \citep{lynch1977heterosynaptic}\citep{chistiakova2014heterosynaptic}.\par

\noindent A strong input form the presynaptic main excitatory pathway can be cumulated via spatial or temporal summation to surpass a threshold that ignites an ictal activity. Upon receiving a robust input, the activity-dependent depressed positive feedbacks are fully activated, due to \(Ca^{++}\) rush, and the accumulated vesicles fuse into the presynaptic membrane to release an excessive amount of excitatory neurotransmitters into the competition zone. The positive feedback excitatory loops are unstable and collapse spontaneously; yet the excitatory neurotransmitters released into the competition zone render the presynaptic feedback or feedforward inhibitory synapse(s) functionally obtunded. Additional molecular mechanism can also contribute to sudden inhibitory stupor within the competition zone; excessive unstable positive feedbacks that ignite the ictal behavior cause the inhibitory feedbacks, not the inhibitory feedforwards, to oscillate rapidly. Intensely activated, dendritic \(GABA_A\) receptors excite rather than inhibit the postsynaptic membrane \citep{staley1995ionic} of the principal neuron(s) with the competition zone.\par

\noindent The molecular mechanism of GABA-mediated membrane depolarization involves differential anionic concentration shift during intense \(GABA_A\) receptor activation. The membrane potential is positive to the \(Cl^-\) reversal potential \citep{bormann1987mechanism}, so that the \(Cl^-\) ions flux is inward and the intracellular \(Cl^-\) concentration increases, but the membrane potential is negative to the \(HCO_3^-\) reversal potential \citep{huguenard1986whole}, so that the intracellular \(HCO_3^-\) concentration decreases due to the efflux of \(HCO_3^-\) ions. The decrease in the intracellular \(HCO_3^-\) concentration, drives the diffusion of \(CO_2\) across the dendritic membrane, allowing continual regeneration of intracellular \(HCO_3^-\) ions by carbonic anhydrase at a rate that exceeds the removal of intracellular \(Cl^-\) ions \citep{staley1994role}. Accordingly; The electrochemical gradient for \(Cl^-\) collapses more significantly than does the \(HCO_3^-\) gradient, producing a shift in the \(GABA_A\) reversal potential toward the \(HCO_3^-\) reversal potential that is sufficient to explain the depolarizing response.\par

\noindent Due to the functional stupor of the inhibitory pathways; the presynaptic main excitatory input is released to simultaneously activate the postsynaptic principle neurons forming an escape rhythm. The action potential generated by uninhibited ignition tends to have echoic reverberations along the neuronal membrane. NMDA receptors are very sensitive to changes in the membrane potential; wherein extracellular \(Mg^{++}\) and \(Zn^{++}\) ions bind to specific sites on the receptor, blocking the passage of other cations through the ion channel. Reverberating depolarization of the neuronal membrane, mainly the somatic and dendritic membrane, dislodges and repels the \(Mg^{++}\) and \(Zn^{++}\) ions from the pore, thus allowing a voltage-dependent flow of \(Ca^{++}\) and \(Na^+\) ions into the cell and \(K^+\) out of the cell \citep{cull2001nmda}\citep{liu2000recent}. The influx of cations creates a local signal that can; in the absence of anion-dependent inhibitory control which, besides type-A \(K^+\) channels, prevent reverberating depolarization; retrigger \(Na^+\) voltage-gated ion channels. Such behavior enables multiple waves of \(Na^+\) influx, creating a train of repetitive spikes which is followed by a subsequent hyperpolarization generated by \(Ca^{++}\)-dependent \(K^+\) channels. This response is the paroxysmal depolarization shift, which is a hallmark of epilepsy on the cellular level.

\subsection*{MOLECULAR MECHANISM OF PROPAGATION AND TERMINATION:}

If the number of principal neurons of an ictogenic focus is not large enough to overcome the zone of inhibition exerted onto their axon terminals, the bursting activity would not be electrochemically transmitted via synaptic communication to the connected remote brain loci. Additionally; the synchronous activity of said small number of neurons would not significantly disturb the extraneuronal ionic environment, thus a trigger wave would not be created leading to failure of recruitment of adjacent non-connected neurons of nearby loci. However; when the number of principal neurons of an ictogenic focus is large enough, the bursting activity makes its way through the confinement and tends to propagate. Said large number of neurons will likely overcome the zone of inhibition and transmit the ictal activity via synaptic communication to the connected remote loci.\par

\noindent Besides; synchronous activity of a large number of principal neurons within an ictal focus can potentially disturb the extraneuronal ionic environment, generating a trigger wave which recruits adjacent non-connected neurons of the nearby neural networks. Ictal activity, of a large number of neurons within an ictogenic focus, induces regional elevation of extracellular \(K^+\) ions due to persistent efflux. As mentioned before; neuronal resting membrane potential is very sensitive to changes in extracellular \(K^+\) concentration. Regional rise in the levels of extracellular \(K^+\) ions reduces the magnitude of potassium gradient across the cell membrane of the adjacent neurons; and therefore, shifts the absolute value of the resting membrane potential to depolarization. The adjacent neurons become hyperexcitable and respond by producing an ictal activity. The process of recruitment is self-reinforced generating a trigger wave that promulgates spatially with a defined speed. Active synaptic clearance brings about rebalance among the opposed neurotransmitters within the competition zone, so that the inhibitory control reimposes its effect and decrease the probability of firing of the principle neurons within the ictal circuit demolishing the synchronous escape rhythm; consequently the ictal episode is terminated. Reactive inhibitory reimposition contributes to the altered state of consciousness after an epileptic seizure.

\subsection*{EFFECT OF ANTIEPILEPTICS ON THE COMPETITION ZONE:}

The competition zone is a functional, not an anatomical, entity; wherein the mechanism of action of antiepileptics can be applied to. This subsection discusses the effects of antiepileptics on the competition zone because of the paradox that results from a molecular mechanism that is believed to contribute to the functional stupor of the inhibitory synapse(s), which results from the excessive unstable positive feedbacks that ignite an ictal behavior. The mechanism implicates that an intensely activated, dendritic \(GABA_A\) receptors excite rather than inhibit the postsynaptic membrane \citep{staley1995ionic}. Accordingly; drugs that enhance the effect of the neurotransmitter GABA at the \(GABA_A\) receptors, as benzodiazepines, can effectively prevent seizures as they prevent the main input summation, so they prevent the ignition of the activity-dependent positive feedback excitatory loops in the first place. But; if the main input bypasses such inhibitory effect and ignites the unstable positive feedbacks, the overstimulated \(GABA_A\) receptor, by the drug, would enhance the excitation of the postsynaptic membrane of the principal neuron(s) and functionally contribute, and further prolong, the presynaptic inhibitory stupor.\par

\noindent Consequently; drugs that enhance the effect of the neurotransmitter GABA at the \(GABA_A\) receptors, as benzodiazepines and the like, exacerbate epileptic seizures that develop during using them as a preventive anticonvulsant. Whilst paradoxically; such medications are more effective if they are used to terminate an active attack as they boost the effect of inhibitory synapse(s) on the stable escape rhythm; which results after spontaneous collapse of the unstable excitatory positive feedbacks. Besides, they inhibit the electrochemical propagation of the ictal activity by consolidating the integrity of the inhibitory zone on the synaptic terminals of the ictogenic principal neurons of the primary ictal focus.

\section*{CONCLUSION:}
In short; chronic epilepsy is a disease of mal-architecture of the neural circuits that results from genetically-predisposed, defective wiring in the spontaneous type; or neural dysgenesis in the acquired type. Said mal-architecture leads to formation of activity-dependent excitatory positive feedbacks that can ignite an ictal activity in response to robust stimulation. Such ictal activity tends to propagate electrochemically via synaptic communication to the connected distant loci, and chemically via a trigger wave to the adjacent non-connected loci. The ictal activity is terminated upon reimposition of the functionally obtunded inhibitory control over the competition zone within the primary ictal focus.


\newpage
\thispagestyle{empty}
\bibliographystyle{humannat}
\bibliography{myref}

\begin{thebibliography}{}

\bibitem[\protect\astroncite{Allen et~al.}{2013}]{allen2013novo}
Allen, A.~S., S.~F. Berkovic, P.~Cossette, N.~Delanty, D.~Dlugos, E.~E.
  Eichler, M.~P. Epstein, T.~Glauser, D.~B. Goldstein, Y.~Han,
  et~al.\leavevmode\nopagebreak\newline 2013.
\newblock De novo mutations in epileptic encephalopathies.
\newblock {\em Nature}, 501(7466):217.

\bibitem[\protect\astroncite{Allison and
  Pratt}{2003}]{allison2003neuroadaptive}
Allison, C. and J.~Pratt\leavevmode\nopagebreak\newline 2003.
\newblock Neuroadaptive processes in gabaergic and glutamatergic systems in
  benzodiazepine dependence.
\newblock {\em Pharmacology \& therapeutics}, 98(2):171--195.

\bibitem[\protect\astroncite{Blumenfeld}{2005}]{blumenfeld2005cellular}
Blumenfeld, H.\leavevmode\nopagebreak\newline 2005.
\newblock Cellular and network mechanisms of spike-wave seizures.
\newblock {\em Epilepsia}, 46:21--33.

\bibitem[\protect\astroncite{Boison}{2011}]{boison2011methylxanthines}
Boison, D.\leavevmode\nopagebreak\newline 2011.
\newblock Methylxanthines, seizures, and excitotoxicity.
\newblock In {\em Methylxanthines}, Pp.~ 251--266.
\newblock Springer.

\bibitem[\protect\astroncite{Bormann et~al.}{1987}]{bormann1987mechanism}
Bormann, J., O.~P. Hamill, and B.~Sakmann\leavevmode\nopagebreak\newline 1987.
\newblock Mechanism of anion permeation through channels gated by glycine and
  gamma-aminobutyric acid in mouse cultured spinal neurones.
\newblock {\em The Journal of physiology}, 385(1):243--286.

\bibitem[\protect\astroncite{Bracci et~al.}{2001}]{bracci2001dynamic}
Bracci, E., M.~Vreugdenhil, S.~P. Hack, and J.~G.
  Jefferys\leavevmode\nopagebreak\newline 2001.
\newblock Dynamic modulation of excitation and inhibition during stimulation at
  gamma and beta frequencies in the ca1 hippocampal region.
\newblock {\em Journal of Neurophysiology}, 85(6):2412--2422.

\bibitem[\protect\astroncite{Bromfield et~al.}{2006}]{bromfield2006basic}
Bromfield, E.~B., J.~E. Cavazos, and J.~I.
  Sirven\leavevmode\nopagebreak\newline 2006.
\newblock Basic mechanisms underlying seizures and epilepsy.

\bibitem[\protect\astroncite{Buckmaster}{2014}]{buckmaster2014does}
Buckmaster, P.~S.\leavevmode\nopagebreak\newline 2014.
\newblock Does mossy fiber sprouting give rise to the epileptic state?
\newblock In {\em Issues in Clinical Epileptology: A View from the Bench}, Pp.~
  161--168.
\newblock Springer.

\bibitem[\protect\astroncite{Carvill et~al.}{2014}]{carvill2014gabra1}
Carvill, G.~L., S.~Weckhuysen, J.~M. McMahon, C.~Hartmann, R.~S. M{\o}ller,
  H.~Hjalgrim, J.~Cook, E.~Geraghty, B.~J. O’roak, S.~Petrou,
  et~al.\leavevmode\nopagebreak\newline 2014.
\newblock Gabra1 and stxbp1: novel genetic causes of dravet syndrome.
\newblock {\em Neurology}, 82(14):1245--1253.

\bibitem[\protect\astroncite{Chistiakova
  et~al.}{2014}]{chistiakova2014heterosynaptic}
Chistiakova, M., N.~M. Bannon, M.~Bazhenov, and
  M.~Volgushev\leavevmode\nopagebreak\newline 2014.
\newblock Heterosynaptic plasticity: multiple mechanisms and multiple roles.
\newblock {\em The Neuroscientist}, 20(5):483--498.

\bibitem[\protect\astroncite{Cull-Candy et~al.}{2001}]{cull2001nmda}
Cull-Candy, S., S.~Brickley, and M.~Farrant\leavevmode\nopagebreak\newline
  2001.
\newblock Nmda receptor subunits: diversity, development and disease.
\newblock {\em Current opinion in neurobiology}, 11(3):327--335.

\bibitem[\protect\astroncite{Da~Silva et~al.}{2003}]{da2003epilepsies}
Da~Silva, F.~L., W.~Blanes, S.~N. Kalitzin, J.~Parra, P.~Suffczynski, and D.~N.
  Velis\leavevmode\nopagebreak\newline 2003.
\newblock Epilepsies as dynamical diseases of brain systems: basic models of
  the transition between normal and epileptic activity.
\newblock {\em Epilepsia}, 44:72--83.

\bibitem[\protect\astroncite{Debanne et~al.}{1994}]{debanne1994asynchronous}
Debanne, D., B.~G{\"a}hwiler, and S.~M. Thompson\leavevmode\nopagebreak\newline
  1994.
\newblock Asynchronous pre-and postsynaptic activity induces associative
  long-term depression in area ca1 of the rat hippocampus in vitro.
\newblock {\em Proceedings of the National Academy of Sciences},
  91(3):1148--1152.

\bibitem[\protect\astroncite{Fisher et~al.}{2005}]{fisher2005epileptic}
Fisher, R.~S., W.~V.~E. Boas, W.~Blume, C.~Elger, P.~Genton, P.~Lee, and
  J.~Engel~Jr\leavevmode\nopagebreak\newline 2005.
\newblock Epileptic seizures and epilepsy: definitions proposed by the
  international league against epilepsy (ilae) and the international bureau for
  epilepsy (ibe).
\newblock {\em Epilepsia}, 46(4):470--472.

\bibitem[\protect\astroncite{Frank et~al.}{2006}]{frank2006reduced}
Frank, H.~Y., M.~Mantegazza, R.~E. Westenbroek, C.~A. Robbins, F.~Kalume, K.~A.
  Burton, W.~J. Spain, G.~S. McKnight, T.~Scheuer, and W.~A.
  Catterall\leavevmode\nopagebreak\newline 2006.
\newblock Reduced sodium current in gabaergic interneurons in a mouse model of
  severe myoclonic epilepsy in infancy.
\newblock {\em Nature neuroscience}, 9(9):1142.

\bibitem[\protect\astroncite{G{\"o}tz-Trabert et~al.}{2008}]{gotz2008spread}
G{\"o}tz-Trabert, K., C.~Hauck, K.~Wagner, S.~Fauser, and
  A.~Schulze-Bonhage\leavevmode\nopagebreak\newline 2008.
\newblock Spread of ictal activity in focal epilepsy.
\newblock {\em Epilepsia}, 49(9):1594--1601.

\bibitem[\protect\astroncite{Huguenard and Alger}{1986}]{huguenard1986whole}
Huguenard, J. and B.~Alger\leavevmode\nopagebreak\newline 1986.
\newblock Whole-cell voltage-clamp study of the fading of gaba-activated
  currents in acutely dissociated hippocampal neurons.
\newblock {\em Journal of Neurophysiology}, 56(1):1--18.

\bibitem[\protect\astroncite{Jett}{2012}]{jett2012chemical}
Jett, D.~A.\leavevmode\nopagebreak\newline 2012.
\newblock Chemical toxins that cause seizures.
\newblock {\em Neurotoxicology}, 33(6):1473--1475.

\bibitem[\protect\astroncite{Khan and Al~Baradie}{2012}]{khan2012epileptic}
Khan, S. and R.~Al~Baradie\leavevmode\nopagebreak\newline 2012.
\newblock Epileptic encephalopathies: an overview.
\newblock {\em Epilepsy research and treatment}, 2012.

\bibitem[\protect\astroncite{Liu and Zhang}{2000}]{liu2000recent}
Liu, Y. and J.~Zhang\leavevmode\nopagebreak\newline 2000.
\newblock Recent development in nmda receptors.
\newblock {\em Chinese medical journal}, 113(10):948--956.

\bibitem[\protect\astroncite{Lynch et~al.}{1977}]{lynch1977heterosynaptic}
Lynch, G.~S., T.~Dunwiddie, and V.~Gribkoff\leavevmode\nopagebreak\newline
  1977.
\newblock Heterosynaptic depression: a postsynaptic correlate of long-term
  potentiation.
\newblock {\em Nature}, 266(5604):737.

\bibitem[\protect\astroncite{Macdonald and Kang}{2012}]{macdonald2012mrna}
Macdonald, R.~L. and J.-Q. Kang\leavevmode\nopagebreak\newline 2012.
\newblock mrna surveillance and endoplasmic reticulum quality control processes
  alter biogenesis of mutant gabaa receptor subunits associated with genetic
  epilepsies.
\newblock {\em Epilepsia}, 53:59--70.

\bibitem[\protect\astroncite{Margineanu}{2010}]{margineanu2010epileptic}
Margineanu, D.~G.\leavevmode\nopagebreak\newline 2010.
\newblock Epileptic hypersynchrony revisited.
\newblock {\em Neuroreport}, 21(15):963--967.

\bibitem[\protect\astroncite{McPhee and
  Hammer}{1995}]{mcphee1995pathophysiology}
McPhee, S.~J. and G.~D. Hammer\leavevmode\nopagebreak\newline 1995.
\newblock {\em Pathophysiology of Disease: An Introduction to Clinical
  Medicine, (Lange Medical Books)}.
\newblock USA.

\bibitem[\protect\astroncite{Misulis and Head}{2003}]{misulis2003essentials}
Misulis, K.~E. and T.~C. Head\leavevmode\nopagebreak\newline 2003.
\newblock {\em Essentials of clinical neurophysiology}, volume~1.
\newblock Garland Science.

\bibitem[\protect\astroncite{Moran et~al.}{2004}]{moran2004epilepsy}
Moran, N., K.~Poole, G.~Bell, J.~Solomon, S.~Kendall, M.~McCarthy,
  D.~McCormick, L.~Nashef, J.~Sander, and
  S.~Shorvon\leavevmode\nopagebreak\newline 2004.
\newblock Epilepsy in the united kingdom: seizure frequency and severity,
  anti-epileptic drug utilization and impact on life in 1652 people with
  epilepsy.
\newblock {\em Seizure}, 13(6):425--433.

\bibitem[\protect\astroncite{Noebels et~al.}{2012}]{noebels2012jasper}
Noebels, J., M.~Avoli, M.~Rogawski, R.~Olsen, and
  A.~Delgado-Escueta\leavevmode\nopagebreak\newline 2012.
\newblock {\em Jasper's Basic Mechanisms of the Epilepsies}.
\newblock Oxford University Press.

\bibitem[\protect\astroncite{Purves et~al.}{2001}]{purves2001neuroscience}
Purves, D., G.~Augustine, D.~Fitzpatrick, L.~Katz, A.~LaMantia, J.~McNamara,
  and S.~Williams\leavevmode\nopagebreak\newline 2001.
\newblock Neuroscience 2nd edition. sunderland (ma) sinauer associates.

\bibitem[\protect\astroncite{Ran et~al.}{2014}]{ran2014epilepsygene}
Ran, X., J.~Li, Q.~Shao, H.~Chen, Z.~Lin, Z.~S. Sun, and
  J.~Wu\leavevmode\nopagebreak\newline 2014.
\newblock Epilepsygene: a genetic resource for genes and mutations related to
  epilepsy.
\newblock {\em Nucleic acids research}, 43(D1):D893--D899.

\bibitem[\protect\astroncite{Scharfman}{2007}]{scharfman2007neurobiology}
Scharfman, H.~E.\leavevmode\nopagebreak\newline 2007.
\newblock The neurobiology of epilepsy.
\newblock {\em Current neurology and neuroscience reports}, 7(4):348--354.

\bibitem[\protect\astroncite{Schulze-Bonhage and
  K{\"u}hn}{2008}]{schulze2008unpredictability}
Schulze-Bonhage, A. and A.~K{\"u}hn\leavevmode\nopagebreak\newline 2008.
\newblock Unpredictability of seizures and the burden of epilepsy.
\newblock {\em Seizure prediction in epilepsy: from basic mechanisms to
  clinical applications}.

\bibitem[\protect\astroncite{Song et~al.}{2000}]{song2000competitive}
Song, S., K.~D. Miller, and L.~F. Abbott\leavevmode\nopagebreak\newline 2000.
\newblock Competitive hebbian learning through spike-timing-dependent synaptic
  plasticity.
\newblock {\em Nature neuroscience}, 3(9):919.

\bibitem[\protect\astroncite{Staley}{1994}]{staley1994role}
Staley, K.\leavevmode\nopagebreak\newline 1994.
\newblock The role of an inwardly rectifying chloride conductance in
  postsynaptic inhibition.
\newblock {\em Journal of neurophysiology}, 72(1):273--284.

\bibitem[\protect\astroncite{Staley}{2015}]{staley2015molecular}
Staley, K.\leavevmode\nopagebreak\newline 2015.
\newblock Molecular mechanisms of epilepsy.
\newblock {\em Nature neuroscience}, 18(3):367.

\bibitem[\protect\astroncite{Staley et~al.}{1995}]{staley1995ionic}
Staley, K.~J., B.~L. Soldo, and W.~R. Proctor\leavevmode\nopagebreak\newline
  1995.
\newblock Ionic mechanisms of neuronal excitation by inhibitory gabaa
  receptors.
\newblock {\em Science}, 269(5226):977--981.

\bibitem[\protect\astroncite{Sutula and Dudek}{2007}]{sutula2007unmasking}
Sutula, T.~P. and F.~E. Dudek\leavevmode\nopagebreak\newline 2007.
\newblock Unmasking recurrent excitation generated by mossy fiber sprouting in
  the epileptic dentate gyrus: an emergent property of a complex system.
\newblock {\em Progress in brain research}, 163:541--563.

\bibitem[\protect\astroncite{Swartz}{2014}]{swartz2014mechanism}
Swartz, C.~M.\leavevmode\nopagebreak\newline 2014.
\newblock A mechanism of seizure induction by electricity and its clinical
  implications.
\newblock {\em The journal of ECT}, 30(2):94--97.

\bibitem[\protect\astroncite{Veeramah et~al.}{2013}]{veeramah2013exome}
Veeramah, K.~R., L.~Johnstone, T.~M. Karafet, D.~Wolf, R.~Sprissler,
  J.~Salogiannis, A.~Barth-Maron, M.~E. Greenberg, T.~Stuhlmann, S.~Weinert,
  et~al.\leavevmode\nopagebreak\newline 2013.
\newblock Exome sequencing reveals new causal mutations in children with
  epileptic encephalopathies.
\newblock {\em Epilepsia}, 54(7):1270--1281.

\bibitem[\protect\astroncite{Wester and McBain}{2014}]{wester2014behavioral}
Wester, J.~C. and C.~J. McBain\leavevmode\nopagebreak\newline 2014.
\newblock Behavioral state-dependent modulation of distinct interneuron
  subtypes and consequences for circuit function.
\newblock {\em Current opinion in neurobiology}, 29:118--125.

\bibitem[\protect\astroncite{Zhang et~al.}{2009}]{zhang2009surviving}
Zhang, W., R.~Yamawaki, X.~Wen, J.~Uhl, J.~Diaz, D.~A. Prince, and P.~S.
  Buckmaster\leavevmode\nopagebreak\newline 2009.
\newblock Surviving hilar somatostatin interneurons enlarge, sprout axons, and
  form new synapses with granule cells in a mouse model of temporal lobe
  epilepsy.
\newblock {\em Journal of Neuroscience}, 29(45):14247--14256.

\end{thebibliography}


@book{mcphee1995pathophysiology,
  title={Pathophysiology of Disease: An Introduction to Clinical Medicine, (Lange Medical Books)},
  author={McPhee, Stephen J and Hammer, Gary D},
  year={1995},
  publisher={USA}
}
@article{bromfield2006basic,
  title={Basic mechanisms underlying seizures and epilepsy},
  author={Bromfield, Edward B and Cavazos, Jos{\'e} E and Sirven, Joseph I},
  year={2006},
  publisher={American Epilepsy Society}
}
@article{blumenfeld2005cellular,
  title={Cellular and network mechanisms of spike-wave seizures},
  author={Blumenfeld, Hal},
  journal={Epilepsia},
  volume={46},
  pages={21--33},
  year={2005},
  publisher={Wiley Online Library}
}
@article{da2003epilepsies,
  title={Epilepsies as dynamical diseases of brain systems: basic models of the transition between normal and epileptic activity},
  author={Da Silva, Fernando Lopes and Blanes, Wouter and Kalitzin, Stiliyan N and Parra, Jaime and Suffczynski, Piotr and Velis, Demetrios N},
  journal={Epilepsia},
  volume={44},
  pages={72--83},
  year={2003},
  publisher={Wiley Online Library}
}
@article{margineanu2010epileptic,
  title={Epileptic hypersynchrony revisited},
  author={Margineanu, Doru Georg},
  journal={Neuroreport},
  volume={21},
  number={15},
  pages={963--967},
  year={2010},
  publisher={LWW}
}
@article{gotz2008spread,
  title={Spread of ictal activity in focal epilepsy},
  author={G{\"o}tz-Trabert, Katrin and Hauck, Christoph and Wagner, Kathrin and Fauser, Susanne and Schulze-Bonhage, Andreas},
  journal={Epilepsia},
  volume={49},
  number={9},
  pages={1594--1601},
  year={2008},
  publisher={Wiley Online Library}
}
@article{fisher2005epileptic,
  title={Epileptic seizures and epilepsy: definitions proposed by the International League Against Epilepsy (ILAE) and the International Bureau for Epilepsy (IBE)},
  author={Fisher, Robert S and Boas, Walter Van Emde and Blume, Warren and Elger, Christian and Genton, Pierre and Lee, Phillip and Engel Jr, Jerome},
  journal={Epilepsia},
  volume={46},
  number={4},
  pages={470--472},
  year={2005},
  publisher={Wiley Online Library}
}
@book{misulis2003essentials,
  title={Essentials of clinical neurophysiology},
  author={Misulis, Karl E and Head, Thomas Channing},
  volume={1},
  year={2003},
  publisher={Garland Science}
}
@book{noebels2012jasper,
  title={Jasper's Basic Mechanisms of the Epilepsies},
  author={Noebels, Jeffrey and Avoli, Massimo and Rogawski, Michael and Olsen, Richard and Delgado-Escueta, Antonio},
  year={2012},
  publisher={Oxford University Press}
}
@article{schulze2008unpredictability,
  title={Unpredictability of Seizures and the Burden of Epilepsy},
  author={Schulze-Bonhage, Andreas and K{\"u}hn, Anne},
  journal={Seizure prediction in epilepsy: from basic mechanisms to clinical applications},
  year={2008},
  publisher={John Wiley \& Sons}
}
@article{bracci2001dynamic,
  title={Dynamic modulation of excitation and inhibition during stimulation at gamma and beta frequencies in the CA1 hippocampal region},
  author={Bracci, Enrico and Vreugdenhil, Martin and Hack, Stephen P and Jefferys, John GR},
  journal={Journal of Neurophysiology},
  volume={85},
  number={6},
  pages={2412--2422},
  year={2001},
  publisher={American Physiological Society Bethesda, MD}
}
@article{wester2014behavioral,
  title={Behavioral state-dependent modulation of distinct interneuron subtypes and consequences for circuit function},
  author={Wester, Jason C and McBain, Chris J},
  journal={Current opinion in neurobiology},
  volume={29},
  pages={118--125},
  year={2014},
  publisher={Elsevier}
}
@article{staley2015molecular,
  title={Molecular mechanisms of epilepsy},
  author={Staley, Kevin},
  journal={Nature neuroscience},
  volume={18},
  number={3},
  pages={367},
  year={2015},
  publisher={Nature Publishing Group}
}
@article{swartz2014mechanism,
  title={A mechanism of seizure induction by electricity and its clinical implications},
  author={Swartz, Conrad M},
  journal={The journal of ECT},
  volume={30},
  number={2},
  pages={94--97},
  year={2014},
  publisher={LWW}
}
@article{jett2012chemical,
  title={Chemical toxins that cause seizures},
  author={Jett, David A},
  journal={Neurotoxicology},
  volume={33},
  number={6},
  pages={1473--1475},
  year={2012},
  publisher={Elsevier}
}
@incollection{boison2011methylxanthines,
  title={Methylxanthines, seizures, and excitotoxicity},
  author={Boison, Detlev},
  booktitle={Methylxanthines},
  pages={251--266},
  year={2011},
  publisher={Springer}
}
@article{allison2003neuroadaptive,
  title={Neuroadaptive processes in GABAergic and glutamatergic systems in benzodiazepine dependence},
  author={Allison, C and Pratt, JA},
  journal={Pharmacology \& therapeutics},
  volume={98},
  number={2},
  pages={171--195},
  year={2003},
  publisher={Elsevier}
}
@article{allen2013novo,
  title={De novo mutations in epileptic encephalopathies},
  author={Allen, Andrew S and Berkovic, Samuel F and Cossette, Patrick and Delanty, Norman and Dlugos, Dennis and Eichler, Evan E and Epstein, Michael P and Glauser, Tracy and Goldstein, David B and Han, Yujun and others},
  journal={Nature},
  volume={501},
  number={7466},
  pages={217},
  year={2013},
  publisher={Nature Publishing Group}
}
@article{veeramah2013exome,
  title={Exome sequencing reveals new causal mutations in children with epileptic encephalopathies},
  author={Veeramah, Krishna R and Johnstone, Laurel and Karafet, Tatiana M and Wolf, Daniel and Sprissler, Ryan and Salogiannis, John and Barth-Maron, Asa and Greenberg, Michael E and Stuhlmann, Till and Weinert, Stefanie and others},
  journal={Epilepsia},
  volume={54},
  number={7},
  pages={1270--1281},
  year={2013},
  publisher={Wiley Online Library}
}
@article{khan2012epileptic,
  title={Epileptic encephalopathies: an overview},
  author={Khan, Sonia and Al Baradie, Raidah},
  journal={Epilepsy research and treatment},
  volume={2012},
  year={2012},
  publisher={Hindawi}
}
@misc{purves2001neuroscience,
  title={Neuroscience 2nd Edition. Sunderland (MA) Sinauer Associates},
  author={Purves, Dale and Augustine, GJ and Fitzpatrick, D and Katz, LC and LaMantia, AS and McNamara, JO and Williams, SM},
  year={2001},
  publisher={Inc}
}
@article{scharfman2007neurobiology,
  title={The neurobiology of epilepsy},
  author={Scharfman, Helen E},
  journal={Current neurology and neuroscience reports},
  volume={7},
  number={4},
  pages={348--354},
  year={2007},
  publisher={Springer}
}
@article{moran2004epilepsy,
  title={Epilepsy in the United Kingdom: seizure frequency and severity, anti-epileptic drug utilization and impact on life in 1652 people with epilepsy},
  author={Moran, NF and Poole, K and Bell, G and Solomon, J and Kendall, S and McCarthy, M and McCormick, D and Nashef, L and Sander, J and Shorvon, SD},
  journal={Seizure},
  volume={13},
  number={6},
  pages={425--433},
  year={2004},
  publisher={Elsevier}
}
@article{macdonald2012mrna,
  title={mRNA surveillance and endoplasmic reticulum quality control processes alter biogenesis of mutant GABAA receptor subunits associated with genetic epilepsies},
  author={Macdonald, Robert L and Kang, Jing-Qiong},
  journal={Epilepsia},
  volume={53},
  pages={59--70},
  year={2012},
  publisher={Wiley Online Library}
}
@article{frank2006reduced,
  title={Reduced sodium current in GABAergic interneurons in a mouse model of severe myoclonic epilepsy in infancy},
  author={Frank, H Yu and Mantegazza, Massimo and Westenbroek, Ruth E and Robbins, Carol A and Kalume, Franck and Burton, Kimberly A and Spain, William J and McKnight, G Stanley and Scheuer, Todd and Catterall, William A},
  journal={Nature neuroscience},
  volume={9},
  number={9},
  pages={1142},
  year={2006},
  publisher={Nature Publishing Group}
}
@article{carvill2014gabra1,
  title={GABRA1 and STXBP1: novel genetic causes of Dravet syndrome},
  author={Carvill, Gemma L and Weckhuysen, Sarah and McMahon, Jacinta M and Hartmann, Corinna and M{\o}ller, Rikke S and Hjalgrim, Helle and Cook, Joseph and Geraghty, Eileen and O’roak, Brian J and Petrou, Steve and others},
  journal={Neurology},
  volume={82},
  number={14},
  pages={1245--1253},
  year={2014},
  publisher={AAN Enterprises}
}
@article{ran2014epilepsygene,
  title={EpilepsyGene: a genetic resource for genes and mutations related to epilepsy},
  author={Ran, Xia and Li, Jinchen and Shao, Qianzhi and Chen, Huiqian and Lin, Zhongdong and Sun, Zhong Sheng and Wu, Jinyu},
  journal={Nucleic acids research},
  volume={43},
  number={D1},
  pages={D893--D899},
  year={2014},
  publisher={Oxford University Press}
}
@article{zhang2009surviving,
  title={Surviving hilar somatostatin interneurons enlarge, sprout axons, and form new synapses with granule cells in a mouse model of temporal lobe epilepsy},
  author={Zhang, Wei and Yamawaki, Ruth and Wen, Xiling and Uhl, Justin and Diaz, Jessica and Prince, David A and Buckmaster, Paul S},
  journal={Journal of Neuroscience},
  volume={29},
  number={45},
  pages={14247--14256},
  year={2009},
  publisher={Soc Neuroscience}
}
@incollection{buckmaster2014does,
  title={Does mossy fiber sprouting give rise to the epileptic state?},
  author={Buckmaster, Paul S},
  booktitle={Issues in Clinical Epileptology: A View from the Bench},
  pages={161--168},
  year={2014},
  publisher={Springer}
}
@article{sutula2007unmasking,
  title={Unmasking recurrent excitation generated by mossy fiber sprouting in the epileptic dentate gyrus: an emergent property of a complex system},
  author={Sutula, Thomas P and Dudek, F Edward},
  journal={Progress in brain research},
  volume={163},
  pages={541--563},
  year={2007},
  publisher={Elsevier}
}
@article{song2000competitive,
  title={Competitive Hebbian learning through spike-timing-dependent synaptic plasticity},
  author={Song, Sen and Miller, Kenneth D and Abbott, Larry F},
  journal={Nature neuroscience},
  volume={3},
  number={9},
  pages={919},
  year={2000},
  publisher={Nature Publishing Group}
}
@article{debanne1994asynchronous,
  title={Asynchronous pre-and postsynaptic activity induces associative long-term depression in area CA1 of the rat hippocampus in vitro.},
  author={Debanne, Dominique and G{\"a}hwiler, BH and Thompson, Scott M},
  journal={Proceedings of the National Academy of Sciences},
  volume={91},
  number={3},
  pages={1148--1152},
  year={1994},
  publisher={National Acad Sciences}
}
@article{lynch1977heterosynaptic,
  title={Heterosynaptic depression: a postsynaptic correlate of long-term potentiation},
  author={Lynch, Gary S and Dunwiddie, Thomas and Gribkoff, Valentin},
  journal={Nature},
  volume={266},
  number={5604},
  pages={737},
  year={1977},
  publisher={Nature Publishing Group}
}
@article{chistiakova2014heterosynaptic,
  title={Heterosynaptic plasticity: multiple mechanisms and multiple roles},
  author={Chistiakova, Marina and Bannon, Nicholas M and Bazhenov, Maxim and Volgushev, Maxim},
  journal={The Neuroscientist},
  volume={20},
  number={5},
  pages={483--498},
  year={2014},
  publisher={Sage Publications Sage CA: Los Angeles, CA}
}
@article{staley1995ionic,
  title={Ionic mechanisms of neuronal excitation by inhibitory GABAA receptors},
  author={Staley, Kevin J and Soldo, Brandi L and Proctor, William R},
  journal={Science},
  volume={269},
  number={5226},
  pages={977--981},
  year={1995},
  publisher={American Association for the Advancement of Science}
}
@article{bormann1987mechanism,
  title={Mechanism of anion permeation through channels gated by glycine and gamma-aminobutyric acid in mouse cultured spinal neurones.},
  author={Bormann, Jochen and Hamill, Owen P and Sakmann, Bert},
  journal={The Journal of physiology},
  volume={385},
  number={1},
  pages={243--286},
  year={1987},
  publisher={Wiley Online Library}
}
@article{huguenard1986whole,
  title={Whole-cell voltage-clamp study of the fading of GABA-activated currents in acutely dissociated hippocampal neurons},
  author={Huguenard, JR and Alger, BE},
  journal={Journal of Neurophysiology},
  volume={56},
  number={1},
  pages={1--18},
  year={1986},
  publisher={American Physiological Society Bethesda, MD}
}
@article{staley1994role,
  title={The role of an inwardly rectifying chloride conductance in postsynaptic inhibition},
  author={Staley, KEVIN},
  journal={Journal of neurophysiology},
  volume={72},
  number={1},
  pages={273--284},
  year={1994}
}
@article{cull2001nmda,
  title={NMDA receptor subunits: diversity, development and disease},
  author={Cull-Candy, Stuart and Brickley, Stephen and Farrant, Mark},
  journal={Current opinion in neurobiology},
  volume={11},
  number={3},
  pages={327--335},
  year={2001},
  publisher={Elsevier}
}
@article{liu2000recent,
  title={Recent development in NMDA receptors.},
  author={Liu, Yun and Zhang, Juntian},
  journal={Chinese medical journal},
  volume={113},
  number={10},
  pages={948--956},
  year={2000}
}
\clearpage                       

\end{document}